\documentstyle[aps,pra,psfig,epsf,epsfig]{revtex}
\begin{document}

\twocolumn
\draft
\title{
Continuous-variable
teleportation improvement by photon subtraction via
conditional measurement
}
\author{T. Opatrn\'{y}$^{1,2}$, G. Kurizki$^{1}$, and
D.--G. Welsch$^{3}$
 }
\address{
$^1$ Department of Chemical Physics, Weizmann Institute of Science,
761~00 Rehovot, Israel \\
$^{2}$ Department of Theoretical
Physics, Palack\'{y} University, Svobody 26, 779~00 Olomouc, Czech Republic \\
$^{3}$ Theoretisch-Physikalisches Institut, Friedrich Schiller University,
Max-Wien Platz 1, 07743 Jena, Germany 
}
\date{\today}

\maketitle
\begin{abstract}

We show that the recently proposed scheme of teleportation of continuous
variables [S.L. Braunstein and H.J. Kimble, Phys. Rev. Lett. {\bf 80}, 869
(1998)] can be improved by a conditional measurement  in the preparation of  
the entangled state shared by the sender and the recipient. The conditional
measurement subtracts  photons  from the original entangled two-mode squeezed
vacuum,   by transmitting each mode through a  low-reflectivity beam splitter
and  performing a joint photon-number measurement on the reflected beams. In
this way the degree of entanglement of the  shared state is increased and so is
the fidelity of the teleported state.

\end{abstract}



\section{Introduction}

The transfer of quantum information between distant nodes, as part of
cryptographic or computing schemes, is hampered by the 
losses and particularly the decoherence
incurred by the communication channel, as well as by the lack of
sources that produce {\em perfectly} entangled states
\cite{Bennett}. 
Various
schemes based
on unitary operations and measurements of redundant variables
\cite{Purif} or filtering \cite{Horod}
have been suggested to overcome these problems.
A notable scheme aimed at improving the entanglement of qubits shared by 
distant nodes is quantum privacy amplification (QPA) \cite{QPA}. We have 
recently suggested a modification to the QPA, based on conditional 
measurement (CM) which is designed to select an optimal subensemble of 
partly correlated qubits
according to criteria that ensure significant improvement in the resulting
entanglement (or fidelity), along with high success probability of CM
\cite{OK99}.

The present paper is motivated by a similar need for improving the
recently 
studied
scheme of teleportation of continuous variables 
\cite{BK98,Furusawa} (see also \cite{Ralph,Loock,Milburn}), in the spirit of
the original Einstein--Podolsky--Rosen idea \cite{EPR}. 
The chances of realizing
such teleportation are limited by the available squeezing of the entangled
two-mode state. 
We show that the scheme 
can be improved by CM modifying the entangled state.
Transmitting each mode of the two-mode squeezed vacuum
through a low-reflectivity beam splitter and detecting photons
in the reflected beams, the transmitted modes are prepared in 
an entangled state which differ from the original one in the photons 
subtracted by the measurement. Teleportation is performed 
if the two detectors simultaneously register photons. 
In this case CM increases the degree of entanglement of the shared
state, and an increased fidelity of the teleported state is observed.

\begin{figure}[htb]
\centerline{ 
\begin{tabular}{cc}
\includegraphics[totalheight=3in,angle=-90]{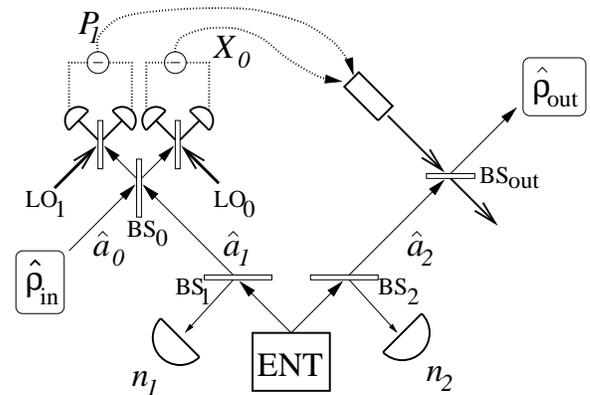}
\end{tabular}
}
\caption{
Teleportation scheme: An input state $\hat \varrho_{\rm in}$ is destroyed by
measurement and it appears, with certain fidelity, at a distant 
node 
as $\hat \varrho_{\rm out}$. The essential means is the entangled state 
created as a two-mode squeezed vacuum in the box ENT. The degree of 
entanglement is improved by CM of the numbers of photons 
$n_1$ and $n_2$ reflected at the 
beam-splitters BS$_1$ and BS$_2$. After mixing the input mode prepared
in the state $\hat \varrho_{\rm in}$ with mode $1$ of the entangled state,
the quadrature components $\hat{x}_0$ and $\hat{p}_1$ are measured and 
their values $X_0$ and $P_1$ are communicated by Alice to Bob via
classical channels (dotted lines). Using these values as displacement 
parameters for shifting the quantum state of mode $2$ on the beam 
splitter BS$_{\rm out}$, Bob creates the output state
$\hat \varrho_{\rm out}$ which imitates $\hat \varrho_{\rm in}$.
}
\label{fschem}
\end{figure}

Let us note that the approach to purification studied in  
\cite{Parker} for continuous variables -- an approach analogous 
to that for spin variables in \cite{Bennett,Purif,QPA} -- does 
not apply to Gaussian entangled states (i.e. states whose 
Wigner functions are Gaussians).
To see this, we recall that the approach uses
beam-splitter transformations 
in combination with quadrature-component measurements. 
The
beam splitter transformations can be
represented by rotations in 
the multi-mode phase space,
and each
quadrature measurement corresponds to a partial integration over a
two-dimensional subspace.
Hence when the original state is Gaussian, then  
rotations and projections of the state ellipsoid onto lower 
dimensional ellipsoids must be performed, and 
it is clear that by this
procedure we can never get a narrower ellipsoid 
(which would correspond to a more strongly entangled state)
than the original one.
Basing on different arguments, the same conclusion is
drawn in \cite{Parker} and therefore the attention has been
restricted to non-Gaussian entangled states.
Alternatively, other than quadrature-component measurement
can be tried to be utilized -- an approach to the problem  
which will be used in what follows.


\section{Teleportation scheme}

For teleporting
an unknown quantum state $\hat \varrho_{\rm in}$
of a single-mode optical field from one
node 
to another one, 
the sender (Alice) and the recipient (Bob) must
share a common two-mode entangled  state. Let us first
recall the original teleportation scheme for continuous variables 
as proposed in \cite{BK98}
(see Fig.~\ref{fschem} without the beam-splitters BS$_1$ and BS$_2$). 
The entangled state is a
two-mode squeezed vacuum, which, in the Fock basis, can be written as 
\begin{eqnarray}
 |\psi_{E}\rangle = \sqrt{1-q^2} \sum_{n=0}^{\infty}q^n 
 |n\rangle_{1}|n\rangle_{2} \,,
 \label{e1}
\end{eqnarray}
where the indices $1$ and $2$ refer to the two modes, and $q$, 
\mbox{$0$ $\!<$ $\!q$ $\!<1$}, 
is a parameter quantifying the strength of squeezing. The first
mode is mixed, on a $50$\%/$50$\% beam splitter BS$_0$, with the 
input mode prepared in a state $\hat\varrho_{\rm in}$ Alice
wishes to teleport. 
Homodyne detections are performed on the two output modes of 
the beam splitter BS$_0$ (using the local oscillators LO$_{0}$
and LO$_{1}$) in order to measure the conjugate quadrature components 
$\hat x_{0}$ and $\hat p_{1}$. 
By sending classical information, Alice communicates the measured values 
$X_{0}$ and $P_{1}$ to Bob, who
uses the value $\sqrt{2}(X_{0}$ $\!+$ $\! iP_{1})$
as a displacement parameter for shifting the quantum state of the
second mode of the entangled state. The resulting quantum state 
$\hat \varrho_{\rm out}$ then imitates $\hat \varrho_{\rm in}$. 
The two states become identical, $\hat \varrho_{\rm out}$ 
$\!\to$ $\!\hat \varrho_{\rm in}$, in the limit of infinite squeezing, 
\mbox{$q$ $\!\to$ $\!1$}.


\subsection{Transformation of the states in quadrature representation}

We restrict ourselves, for simplicity, to pure states and  describe the
transformations in the $\hat x_{j}$ quadrature-component representation 
[$\hat x_{j}$ $\!=$
$\!2^{-1/2}(\hat a_{j}$ $\! +$ $\! \hat a_{j}^{\dag})$, 
$\hat p_{j}$ $\!=$
$\!-2^{-1/2}i(\hat a_{j}$ $\! -$ $\! \hat a_{j}^{\dag})$,
\mbox{$j$ $\! =$ $\! 0,1,2$}]. Let the input-state wave function 
be $\psi_{0}(x_{0})$ $\!\equiv$ $\!\psi_{\rm in}(x_{0})$
and  the entangled state have the wave function 
$\psi_{E}(x_{1},x_{2})$, so that the initial 
overall
wave function is  
\begin{eqnarray}
 \psi_{I}(x_{0},x_{1},x_{2}) = \psi_{0}(x_{0})\psi_{E}(x_{1},x_{2}) .
\end{eqnarray}
We consider the scheme in Fig.~\ref{fschem} without
the beam splitters BS$_1$ and BS$_2$.
Assuming the beam-splitter BS$_0$ mixes the quadratures
as 
\begin{eqnarray}
\hat  x_{0} & \to & 2^{-1/2}(\hat x_{1} + \hat x_{0}) , \nonumber \\
\hat  x_{1} & \to & 2^{-1/2}(\hat x_{1} - \hat x_{0}) , 
\end{eqnarray}
the transformed wave function is 
\begin{eqnarray}
 \psi_{II}(x_{0},x_{1},x_{2}) = \psi_{0}\!\left(\frac{x_{1} + x_{0}}{\sqrt{2}}
 \right)
 \psi_{E}\!\left(\frac{x_{1}- x_{0}}{\sqrt{2}},x_{2}\right) .
\end{eqnarray}
Measuring the quadratures $\hat x_{0}$ and $\hat p_{1}$ to 
obtain the values $X_{0}$ and $P_{1}$, the
(unnormalized) wave function of the mode $2$ reads
\begin{eqnarray}
\lefteqn{
 \psi_{X_{0},P_{1}}(x_{2})  
= (2\pi)^{-1/2}
 \int dx_{1} \ e^{-i P_{1}x_{1}} }\nonumber \\ &&
 \hspace*{4ex}\times\,
 \psi_{0}\!\left(\frac{x_{1} + X_{0}}{\sqrt{2}}\right)
 \psi_{E}\!\left(\frac{x_{1}- X_{0}}{\sqrt{2}},x_{2}\right) .
\end{eqnarray}
The probability density of measuring the values $X_{0}$ and $P_{1}$
is given by
\begin{eqnarray}
{\cal P}(X_{0},P_{1}) 
= \int dx_{2} | \psi_{X_{0},P_{1}}(x_{2})|^{2} .
 \label{prob}
\end{eqnarray}
Using the measured values  $X_{0}$ and $P_{1}$ to 
realize a displacement transformation on the mode $2$,    
\begin{eqnarray}
 \hat  x_{2} & \to & \hat x_{2} - \sqrt{2} X_{0} , \nonumber \\
 \hat  p_{2} & \to & \hat p_{2} + \sqrt{2} P_{1} , 
\end{eqnarray}
the resulting (unnormalized) 
wave function of the mode is found to be
\begin{eqnarray}
\lefteqn{
 \psi_{\rm out}(x_{2})  
= (2\pi)^{-1/2}
 \int dx_{1} \ e^{i P_{1}(\sqrt{2}x_{2} - x_{1})} }\nonumber \\ &&
 \hspace*{2ex}\times\,
 \psi_{0}\!\left(\frac{x_{1} + X_{0}}{\sqrt{2}}\right)
 \psi_{E}\!\left(\frac{x_{1}- X_{0}}{\sqrt{2}},x_{2}-\sqrt{2} X_{0}\right) .
 \label{e-transf}
\end{eqnarray}

An infinitely squeezed two-mode vacuum, $q$ $\!\to$ $\!1$ 
in Eq.~(\ref{e1}),
can be described, apart from normalization, by the Dirac delta function,
\begin{eqnarray}
 \psi_{E}(x_{1},x_{2}) \to \delta(x_{1}-x_{2}) .
\end{eqnarray}
It can easily be checked that Eq.~(\ref{e-transf}) then reduces to
\begin{eqnarray}
 \psi_{\rm out}(x_{2}) \to \psi_{0}(x_{2}) ,
\end{eqnarray}
i.e., the input quantum state is perfectly teleported.
It remains the question of how to improve the fidelity of
teleportation when, as in practice, $\psi_{E}(x_{1},x_{2})$ 
is not infitely squeezed.


\subsection{Transformation to the Fock basis}

For the following it will be convenient to change over to the 
Fock basis. Let us express the input-state wave function in the form of
\begin{eqnarray}
\psi_{0}(x_{0}) = \sum_{n} a_{n}^{(0)} \varphi_{n}(x_{0}) , 
\label{a-expand}
\end{eqnarray}
with $\varphi_{n}(x_{0})$ being the harmonic oscillator energy eigenfunctions 
\begin{eqnarray}  
  \varphi_{n}(x_{0}) = \left(2^n n! \sqrt{\pi}\right)^{-1/2} 
  e^{-x_{0}^2/2} H_{n}(x_{0}) ,  
\end{eqnarray}
$H_{n}(x_{0})$ being the Hermite polynomial,
and let the entangled state  $\psi_{E}(x_{1},x_{2})$ be given by
\begin{eqnarray}
  \psi_{E}(x_{1},x_{2}) = \sum_{k,l} a_{k,l}^{(E)}
  \varphi_{k}(x_{1}) \varphi_{l}(x_{2}) . 
  \label{entexp}
\end{eqnarray} 
In order to find the coefficients $b_{m}(X_0,P_1)$ of the Fock state 
expansion of the (unnormalized) wave function of the teleported 
quantum state
\begin{eqnarray}
 \psi_{\rm out}(x_{2}) = \sum_{m} b_{m}(X_0,P_1) \varphi_{m}(x_{2}) ,
 \label{b-expand}
\end{eqnarray}
we insert $\psi_{\rm out}(x_{2})$ from Eq.~(\ref{e-transf}) into
Eq.~(\ref{b-expand})
and obtain
\begin{eqnarray}
b_{m}(X_{0},P_{1}) 
= \sum_{n} C_{m,n}(X_{0},P_{1}) a_{n},
 \label{outFock}
\end{eqnarray}
where $C_{m,n}(X_{0},P_{1})$ is given by
\begin{eqnarray}
\lefteqn{
 C_{m,n}(X_{0},P_{1}) = (2\pi)^{-1/2} \sum_{k,l}
 B_{m,k}(X_{0},P_{1}) }\nonumber \\ &&
 \hspace*{18ex}\times\, a_{k,l}^{(E)}  D_{l,n}(X_{0},P_{1}),
\end{eqnarray} 
with
\begin{eqnarray}
\lefteqn{
 B_{m,k}(X_{0},P_{1}) = \int dx_{2}\ e^{i\sqrt{2}P_{1}x_{2}}
 \varphi_{m}(x_{2}) }\nonumber \\ &&
 \hspace*{18ex}\times \, \varphi_{k}\!\left( x_{2} \!-\!\sqrt{2} X_{0}\right) 
 \label{bint}
\end{eqnarray}
and 
\begin{eqnarray}
\lefteqn{
 D_{l,n}(X_{0},P_{1}) = \int dx_{1}\ e^{-iP_{1}x_{1}}
 \varphi_{l}\!\left( \frac{x_{1}\!-\!X_{0}}{\sqrt{2}} \right)
 } \nonumber \\ &&
 \hspace*{18ex}\times\, 
 \varphi_{n}\!\left( \frac{x_{1}\! +\! X_{0}}{\sqrt{2}}\right) .
 \label{dint}
\end{eqnarray}
The integrals in Eqs.~(\ref{bint}) and (\ref{dint}) can be expressed in 
a closed form yielding
\begin{eqnarray}
 B_{m,k}(X_{0},P_{1}) = \sqrt{2^{k-m}}\sqrt{\frac{m!}{k!}}
 \left( -\frac{X_{0}\!-\!iP_{1}}{\sqrt{2}} \right)^{k\!-\!m}
 \nonumber \\
 \times \exp \left( -\frac{X_{0}^{2}\!+\!P_{1}^{2}}{2} \right)
 L_{m}^{k-m} \left( X_{0}^{2}\!+\!P_{1}^{2} \right)  
\end{eqnarray}
for $k$ $\!\ge$ $\! m$, and 
\begin{eqnarray}
 B_{k,m}(X_{0},P_{1}) = (-1)^{k-m} B_{m,k}^{*}(X_{0},P_{1}),
\end{eqnarray}
and 
\begin{eqnarray}
 D_{l,n}(X_{0},P_{1}) = \sqrt{2} B_{l,n}^{*}(X_{0},P_{1}) ,
\end{eqnarray}
$L_{m}^{\alpha}(y)$ being the (associated) Laguerre polynomial.


\subsection{Probability, fidelity, and averaged state}

Using the expansion (\ref{b-expand}), the probability density 
of measuring $X_0$ and $P_1$, Eq.~(\ref{prob}), reads as
\begin{eqnarray}
 {\cal P}(X_{0},P_{1}) = \sum_{m} \left| b_{m} (X_{0},P_{1})\right|^{2}.
\label{probX0P1} 
\end{eqnarray}
The fidelity of teleportation is defined by the overlap of the 
input quantum state with the 
(normalized) output quantum state. From Eqs.~(\ref{a-expand}), 
(\ref{b-expand}), and (\ref{probX0P1}) it follows that
\begin{eqnarray}
 F(X_{0},P_{1}) = 
 {\cal P}^{-1}(X_{0},P_{1})
 \sum_{n} \left| a_{n}^{*} b_{n} (X_{0},P_{1})\right|^{2}.
\label{fidelity} 
\end{eqnarray}

So far we have considered the output state under the condition of a particular
measurement outcome $(X_{0},P_{1})$. 
When we ignore the outcome 
and many teleportations take place, then the resulting output
state behaves like a mixture with the 
(unnormalized) 
density matrix elements
\begin{eqnarray}
 \varrho_{m,m'} = \int dX_{0} \int dP_{1} \, b_{m}^{*}(X_{0},P_{1})
 b_{m'}(X_{0},P_{1}). 
\end{eqnarray}
The averaged fidelity in the Fock basis is then given by
\begin{eqnarray}
 F & = & \int dX_{0} \ \int dP_{1} F(X_{0},P_{1}) 
 {\cal P}^{-1}(X_{0},P_{1}) \nonumber \\
 & = & \sum_{m,m'} a^{*}_{m}  \varrho_{m,m'} a_{m'} .
 \label{fidelita}
\end{eqnarray}


\section{Improving entanglement by conditional photon-number measurement}

As already mentioned, perfect teleportation 
requires an infinitely squeezed vacuum, which
is, of course, not available.
The fidelity of the teleported state decreases 
with decreasing squeezing. Our objective here is to increase 
the fidelity by improving 
the entanglement properties of the shared state using conditional
photon-number measurements. It has been shown 
that when a single-mode squeezed vacuum 
is transmitted through a beam splitter 
and a photon-number measurement is performed on the reflected beam,  
then a Schr\"{o}dinger-cat-like state is generated \cite{Dakna}.
Even though photons have been subtracted, the mean number of 
photons remaining in the transmitted state has increased.  


\subsection{Conditionally entangled state}

Let us apply this concept of CM to a two-mode squeezed vacuum and assume
that each mode is transmitted through a low-reflectivity beam splitter 
(BS$_j$ in Fig.~\ref{fschem}, $j$ $\!=$ $\!1,2$)
and the numbers of reflected photons $n_{j}$ are detected. 
Each beam splitter BS$_j$ is described by a transformation matrix
\begin{eqnarray}
 T_{j} = \left( 
 \begin{array}{rr}
 t_{j} & r_{j} \\
 -r_{j} & t_{j}
 \end{array}
   \right)
\end{eqnarray}
with real transmittance $t_{j}$ and real reflectance $r_{j}$, 
for simplicity. These matrices act on the 
operators of the input modes.
After detecting $n_{j}$ photons in the reflected modes, the Fock states
$|k_{j}\rangle$ transform as ($n_{j} \le k_{j}$)
\begin{eqnarray}
 |k_{j}\rangle \to (-1)^{n_{j}}\sqrt{\left( 
 \begin{array}{c}
 k_{j} \\ n_{j}
 \end{array}
  \right)} |r_{j}|^{n_{j}} |t_{j}|^{k_{j}-n_{j}} |k_{j}-n_{j} \rangle 
\end{eqnarray}
(for details and more general beam-splitter
transformations, see \cite{Dakna,Dakna1}). 

The expansion coefficients $a_{k,l}^{(E)}$ of the entangled 
two-mode wave function (\ref{entexp}) are then transformed into  
\begin{eqnarray}
\lefteqn{
 a_{k,l}^{(E,\, {\rm new})}
 = (-1)^{n_{1}\!+\!n_{2}} \sqrt{\frac{(k\!+\!n_{1})!(l\!+\!n_{2})!}
 {k!\ l!\ n_{1}!\ n_{2}!}} 
 }\nonumber \\ &&
 \hspace*{15ex}\times \, |r_{1}|^{n_{1}} |r_{2}|^{n_{2}}
 |t_{1}|^{k} |t_{2}|^{l}
  a_{k+n_{1},l+n_{2}}^{(E,\, {\rm old})} .
\label{coefficient}  
\end{eqnarray}
Note that in this form the wave function is not normalized. The sum of 
the squares of moduli of the coefficients $a_{k,l}^{(E,\, {\rm new})}$ 
gives the probability of the measurement results $n_{1}$ and $n_{2}$.
When the original entangled state a the two-mode squeezed vacuum, 
Eq.~(\ref{e1}), i.e., 
\begin{eqnarray}
 a_{k,l}^{(E,\, {\rm old})} = \sqrt{1-q^2} \ q^k \delta _{k,l},
 \label{aeold}
\end{eqnarray}
then the expansion coefficients (\ref{coefficient}) of the new state read
\begin{eqnarray}
 a_{k,l}^{(E,\, {\rm new})} = (-1)^{n_{1}\!+\!n_{2}}
 \frac{\sqrt{1-q^2}\ (k\!+\!n_{1})!}
 {\sqrt{k!\ (k\!+\!n_{1}\!-\!n_{2})!\ n_{1}!\ n_{2}!}}
 \nonumber \\
 \times \, |r_{1}|^{n_{1}} |r_{2}|^{n_{2}}
 |t_{1}|^{k} |t_{2}|^{k+n_{1}-n_{2}} q^{k+n_{1}} 
 \delta_{k+n_{1},l+n_{2}} \,.
 \label{aenew}
\end{eqnarray}
The most important property of this expression is that the polynomial increase
with $k$ can, for small values of $k$, overcome the exponential decrease
$(q|t_{1}t_{2}|)^{k}$ and thus increase the mean number of photons. This is
especially the case when $|t_{j}|$ is close to unity, i.e. large transmittance
of the beam splitters. The price for that is, however, a decrease in the
probability of detecting the photons.


\subsection{Entropy as entanglement measure}

The increase of the degree of entanglement of the shared state 
produced by CM can be quantified by comparing it with the 
original degree of entanglement. Even though there is no unique
definition of a measure of entanglement for {\em mixed\/} states, 
there is a consensus on defining the degree of entanglement $E$ of 
a two-component system prepared in a {\em pure\/} state 
as the von Neumann entropy of a component. The calculation of the 
partial entropies of the state in Eq.~(\ref{entexp}) 
together with either Eq.~(\ref{aeold}) or (\ref{aenew}) is
simple as the traced states are diagonal in the Fock basis.
We derive 
\begin{eqnarray}
E = - \frac{\sum_{k} |a_{k,k\!+\!n_{1}\!-\!n_{2}}^{(E\,{\rm new})}|^2
 \log |a_{k,k\!+\!n_{1}\!-\!n_{2}}^{(E\,{\rm new})}|^2 }
 {\sum_{k} |a_{k,k\!+\!n_{1}\!-\!n_{2}}^{(E\,{\rm new})}|^2} 
\end{eqnarray} 
($k$ $\!+$ $\!n_{1}$ $\!-$ $\!n_{2}$ $\!\ge$ $\!0$; note that
the entropies of the two components are equal to each other).
If the logarithm base is chosen to be equal to 2, then the entanglement 
is measured in bits (or e-bits).

\begin{figure}[htb]
\centerline{ 
\begin{tabular}{cc}
\includegraphics[totalheight=2.8in,angle=-90]{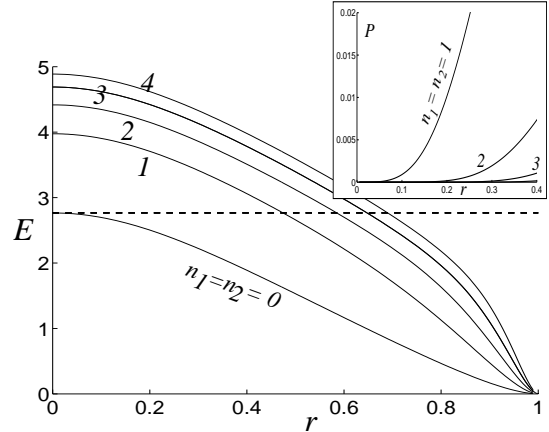}
\end{tabular}
}\caption{ 
Degree of entanglement
(in bits) of the 
two-mode state in Eq.~(\protect\ref{entexp}) together with 
Eq.~(\protect\ref{aenew})
as a function of the reflectance 
$r$ $\!=$ $\!r_{1}$ $\!=$ $\!r_{2}$ 
of the beam splitters BS$_1$ and BS$_2$ in Fig.~\protect\ref{fschem} for
different numbers of detected photons, $n_{1}$ $\!=$ $\!n_{2}$ $\!=$ 
$\!0,1,2,3,4$. The dashed line indicates the 
degree of entanglement 
of the original squeezed vacuum, $q$ $\!=$ $\!0.8178$.
Inset: probability of
detecting $n_{1}$ $\!=$ $\!n_{2}$ $\!=$ $\!1,2,3$ photons 
as a function of the beam-splitter reflectance $r$.
}
\label{fentr1}
\end{figure}

In Fig.~\ref{fentr1} we have shown the dependence of the
degree of entanglement 
and the detection probability on the
beam-splitter reflectance for different numbers of detected photons,
$n_{1}$ $\!=$ $\!n_{2}$. The original squeezed vacuum is chosen such
that $q$ $\!=$ $\!0.8178$, which corresponds to the parameter 
arctanh$\,q$ $\!=$ $\!1.15$ in Ref.~\cite{BK98}. 
We see that after detecting one reflected photon in each 
channel the entanglement can be increased by more than one
bit, and 
the effect increases with the number of detected photons.
However, the probability of detecting more than one
photon may be extremely small. 
Hence, one has to find a compromise between increasing 
degree of entanglement and decreasing success probability.
One should also keep in mind that the partial von Neumann entropy as 
entanglement measure relates to the
maximum information that can be gained about one component
of a two-component system from a measurement on the other component.
Therefore the partial entropy in Fig.~\ref{fentr1} 
represents an upper limit for the quantum communication possibilities 
rather than a direct measure of the quality of 
the teleportation scheme under consideration.

\begin{figure}[htb]
\centerline{ 
\begin{tabular}{cc}
\includegraphics[totalheight=2.8in,angle=-90]{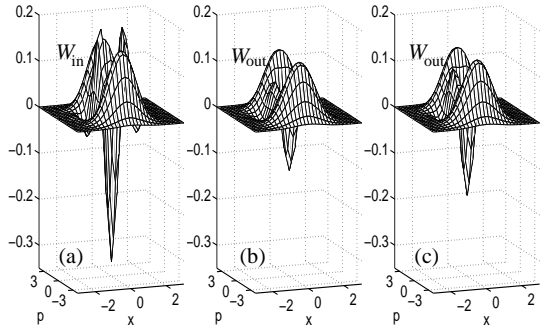}
\end{tabular}
}\caption{ 
(a) Wigner function of the quantum state to be teleported,
$|\Psi\rangle_{\rm in}$ $\!\sim(|\alpha \rangle$ 
$\!-$ $\!|-\alpha \rangle)$, $\alpha$ $\!=$ $\!1.5\,i$; 
(b) Wigner function of the teleported quantum state 
averaged over all measured quadrature-component values
$X_{0}$ and $P_{1}$, the entangled state being the squeezed
vacuum with $q$ $\!=$ $\!0.8178$; 
(c) same as in (b) but for the case 
when the entangled state is the photon-subtracted squeezed vacuum
obtained by CM (\mbox{$n_{1}$ $\! =$ $\! n_{2} $ $\!= 1$}; 
$r$ $\!=$ $\!0.15$).
}
\label{fwig1}
\end{figure}


\section{Results}

We have performed computer simulations in order to test the
method for different input states, especially for
Schr\"{o}dinger cats, which are popular laboratory animals in theoretical
quantum optics (see, e.g., \cite{BK98}). Figure \ref{fwig1}
demonstrates teleportation of the state 
\mbox{$|\Psi\rangle_{\rm in}$ $\!\sim(|\alpha \rangle$ 
$\!-$ $\!|-\alpha \rangle)$} 
with $\alpha$ $\!=$ $\!1.5\,i$, which is chosen to be the same 
as in \cite{BK98}, for comparison. 

\begin{figure}[htb]
\centerline{ 
\begin{tabular}{cc}
\includegraphics[totalheight=2.8in,angle=-90]{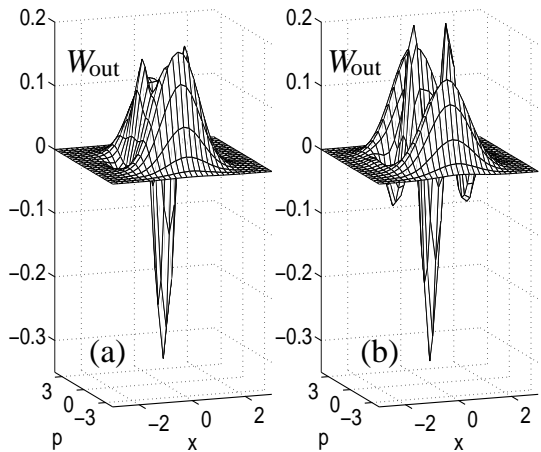}
\end{tabular}
}
\protect\caption{ 
(a) Wigner function of the teleported quantum state  
for $X_{0}$ $\!=$ $\!0.1$ and \mbox{$P_{1}$ $\!=$ $\!0.2$} 
and the input state shown in Fig.~\protect\ref{fwig1}(a) in the case  
when the entangled state is a squeezed vacuum with $q$ $\!=$ $\!0.8178$;
(b) same as in (a) but for the case 
when the entangled state is the photon-subtracted squeezed vacuum
obtained by CM ($n_{1}$ $\! =$ $\! n_{2} $ $\!= 1$;
$r$ $\!=$ $\!0.15$).
}
\label{fwig4}
\end{figure}

\begin{figure}[htb]
\centerline{ 
\begin{tabular}{cc}
\includegraphics[totalheight=2.8in,angle=-90]{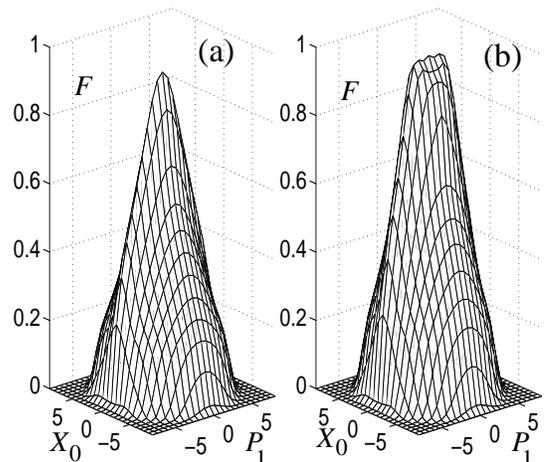} 
\end{tabular}
}
\caption{Fidelity of the teleported quantum state in dependence 
on the measured quadrature-component values $X_{0}$ and $P_{1}$
for the input state shown in Fig.~\protect\ref{fwig1}(a);
(a) the entangled state is a squeezed vacuum ($q$ $\!=$ $\!0.8178$); 
(b) the entangled state is the photon-subtracted squeezed vacuum obtained 
by CM ($n_{1}$ $\! =$ $\! n_{2} $ $\!= 1$;
$r$ $\!=$ $\!0.15$).
}
\label{ffidels}
\end{figure}

\begin{figure}[htb]
\centerline{ 
\begin{tabular}{cc}
\includegraphics[totalheight=2.8in,angle=-90]{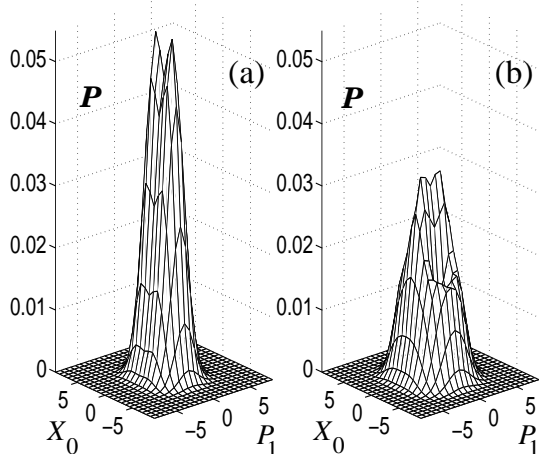}
\end{tabular}
}
\caption{Probability density of measuring the quadrature-component 
values $X_{0}$ and $P_{1}$, Eq.~(\protect\ref{probX0P1}),
for the input state shown in Fig.~\protect\ref{fwig1}(a);
(a) the entangled state is a squeezed vacuum \mbox{($q$ $\!=$ $\!0.8178$)}; 
(b) the entangled state is the photon-subtracted squeezed vacuum
obtained by CM (\mbox{$n_{1}$ $\! =$ $\! n_{2} $ $\!= 1$};
$r$ $\!=$ $\!0.15$).
}
\label{fprobabs}
\end{figure}

The teleported quantum state that is obtained 
for a particular measurement of quadrature-component values
$X_{0}$ and $P_{1}$ is shown in Fig.~\ref{fwig4}.
The results in Figs.~\ref{fwig4}(a) and \ref{fwig4}(b), respectively, 
correspond to the cases when the entangled state is a squeezed 
vacuum ($q$ $\!=$ $\!0.8178$) and a photon-subtracted squeezed 
vacuum with \mbox{$n_{1}$ $\!=$ $\! n_{2}$ $\! =$ $\! 1$},
the probability of producing the state by CM being $0.39$\% 
($r$ $\!=$ $\!0.15$, cf. Fig.~\ref{fentr1}). 
The fidelity of the teleported quantum state, Eq.~(\ref{fidelity}),
is plotted in Fig.~\ref{ffidels} as a function of the measured 
quadrature-component values $X_{0}$ and $P_{1}$, and Fig.~\ref{fprobabs} 
shows the corresponding probability distribution, Eq.~(\ref{probX0P1}). 
We can see that not only the fidelity attains larger values for 
the improved entangled state, Fig.~\ref{ffidels}(b), 
but also the probability distribution is broader for that state, 
Fig.~\ref{fprobabs}(b). The latter is very important. 
If the probability distribution is sharply peaked, then 
Alice actually gains more information about the state, so that there is 
less quantum information to be communicated to Bob.
(Note the extreme case when the ``entangled'' state is simply the vacuum. 
Then Alice measures just the $Q$ function of the state to be ``teleported''. 
The probability distribution of the measured quantities thus carries 
the full information about the state.)

Averaging the fidelity $F(X_0,P_0)$ over the probability distribution 
of the measured quadrature-component values $X_0$ and $P_1$,
we get the averaged fidelity $F$, Eq.~(\ref{fidelita}),
which, for the case in our example, 
attains the value \mbox{$F$ $\!=$ $\!0.6463$}
for the squeezed vacuum and
$F$ $\!=$    
$\! 0.7444$
for the photon-subtracted squeezed vacuum, which is a 
significant increase. The  Wigner function of the averaged  
teleported quantum state is plotted in Fig.~\ref{fwig1}. Again, we 
see that the state in Fig.~\ref{fwig1}(c), which is
teleported by means of the improved entangled state, 
is closer to the original state in Fig.~\ref{fwig1}(a).

\begin{figure}[htb]
\centerline{ 
\begin{tabular}{cc}
\includegraphics[totalheight=2.8in,angle=-90]{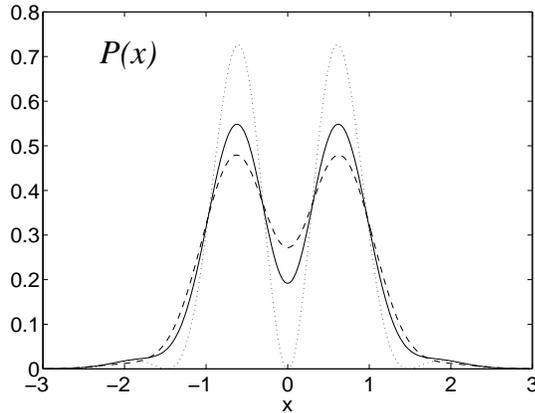}
\end{tabular}
}\caption{ 
Probability distribution of measuring the 
$x$ quadrature in the teleported quantum state 
for the input state shown in Fig.~\protect\ref{fwig1}(a);
dotted line - input state;
dashed line - teleported state if the entangled state is a 
squeezed vacuum \mbox{($q$ $\!=$ $\!0.8178$)}; 
solid line - teleported state if the entangled state is the
photon-subtracted squeezed vacuum 
obtained by CM (\mbox{$n_{1}$ $\! =$ $\! n_{2} $ $\!= 1$};
$r$ $\!=$ $\!0.15$).
}
\label{ffrin}
\end{figure}

The quality of transmission of the intereference fringes 
of the Schr\"{o}dinger cat state can be seen from 
Fig.~\ref{ffrin}, in which the $x$ quadrature distribution of 
the teleported quantum state is plotted.
Whereas the input state shows perfect interference fringes in 
the $x$ quadrature, the teleported states have 
the fringes smeared and their visibilities decreased.
In the example under study, the fringe visibility of the
teleported state is $26.6$\% for the squeezed vacuum and 
$48.2$\% for the photon-subtracted squeezed vacuum
obtained by CM. 

\section{Conclusion}

The results show that conditional photon-number measurement 
can improve the fidelity of teleportation of continuous variables.
With regard to experimental implementation, 
highly efficient single-photon counting is required. 
Even though such counting is at present not as efficient as 
intensity-proportional photodetection, progress has been very fast 
(as it is illustrated by the $88$\% efficiency achieved recently 
\cite{Takeuchi}).
Therefore a realization of the scheme in the near future
may be technically feasible. 

The scheme can also be extended to other types of conditional
measurement. For example, combining  (at the beam splitters BS$_1$ 
and BS$_2$ in Fig.~\ref{fschem}) the modes of the entangled 
two-mode squeezed vacuum with modes prepared in photon-number states, 
zero-photon measurement on the reflected beams then prepares a 
photon-added conditional state. Whereas the 
nonclassical features of a single-mode squeezed vacuum can be 
strongly influenced in this way \cite{Dakna1,added}, 
we have not found a substantial 
improvement of the degree of entanglement of the two 
mode-squeezed vacuum.

We have considered the case when Alice does not know the quantum state
she wishes to teleport. Of course, the quantum communication scheme
can also be applied
to other situations,
e.g., in quantum cryptography or state preparation in a distant place, 
where Alice can know the state. In particular,
Alice can take advantage of her knowledge of the
dependence of the teleportation fidelity on the measured 
quadrature-component values 
in Fig.~\ref{ffidels}
and communicate only the results of measurement which
guarantee high fidelity. In that case,
the teleportation can be regarded as being conditioned not
only by the measured photon numbers in the entangled-state
prepration but also by the measured quadrature-component values. 
This together with 
the possibility of optimization 
of probabilities versus fidelities suggests that there is a rich
area of possible exploration of the
scheme.

\section*{Acknowledgments}

This work was supported by EU (TMR), ISF, Minerva grants, and
the Deutsche Forschungsgemeinschaft.


\end{document}